# Hydrodynamic studies of aqueous two-phase systems in millichannels

Vamsi Vikram Gande, Hima Nandini K, Jagadeesh Korukonda, S Pushpavanam*
Department of Chemical Engineering,
Indian Institute of Technology Madras-600036
*Corresponding author email: spush@iitm.ac.in

## Abstract:

Liquid-liquid segmented flows in microchannels have been extensively investigated in the context of nanoparticle synthesis. The enhanced mixing in the slugs results in monodispersed particles. Earlier studies have focused on Organic-Aqueous Systems (OAS). The nanoparticles synthesized in the presence of organic solutions have limited applications. An alternative green route for the synthesis can be developed using an Aqueous Two-Phase System (ATPS). These systems are characterized by interfacial tensions, which are two orders of magnitude lower than typical organic aqueous systems. In this work, flow patterns and hydrodynamics of ATPS are investigated as a first step. Polyethylene glycol -trisodium citrate system was chosen as ATPS. The objective of this work is to see if any new physics arises in an ATPS system. The low interfacial tension results in high Capillary numbers ($Ca >> 3$) in a microfluidic system. Consequently, the flow observed here is parallel or core-annular. However, in a millichannel, the capillary number becomes lower ($Ca << 1$) for an ATPS system. In this work, experiments were carried out in a millichannel to span different flow patterns. The pattern formation was analyzed and classified into three categories, i.e., slug flow (interfacial tension dominated), transition flow, and core annular flow (inertia dominated). Flow regime maps based on the Reynolds number, Capillary number, and Weber number of each phase were found to be qualitatively similar to those of OAS. Simulations were performed for various interfacial tension values. An interfacial tension value of $1.25 \times 10^{-4}$ N/m was found to yield slug sizes which fitted well with the experimental data. Film thickness was measured experimentally and with simulations compared favorably with the correlations available in the literature for OAS.

## 1. Introduction:

In the past decade, microfluidic systems have been extensively used for material synthesis and various applications in biotechnology [1,2]. The flow in these systems is laminar as the characteristic length scales involved are small. Here due to axial dispersion effects in the laminar regime, the size distribution of nanoparticles obtained at the end of the channel is broad. Several studies have focused on modifying the fluid mechanics in the microchannel to synthesize monodispersed particles. Focus has been on using multiphase segmented flows for enhanced mixing resulting in monodispersed particles [3–6]. Microfluidic systems are robust and efficient for producing uniform droplets (or) slugs with low polydispersity [1]. Several studies have focused on nanoparticle synthesis using segmented flow by introducing an organic phase [3–6]. The enhanced mixing in slugs has gained attention in flow chemistry.

In these systems, the dominant forces are the interfacial forces and viscous forces [1]. At these length scales, the flow is characterized by the formation of droplets or slugs and fluid threads (jets). The flow pattern obtained is affected by flow rates of the fluid phases, fluid viscosities, densities, interfacial tension, and device geometry. Each droplet can be visualized to act as a batch reactor with good internal mixing. In these miniature reactors, mixing is enhanced due to internal vortex circulations and chaotic advection [3,7]. Each droplet or slug moves along the tube length identically. Consequently, the system has a narrow residence time distribution and behaves analogously to a plug flow reactor [8]. Slug flow or segmented flow also has advantages of reduced Taylor Aris dispersion and a large interfacial area [9]. Consequently, they are preferred for generating monodispersed particles.

The use of multiphase flow by the introduction of an immiscible phase (organic phase) has also been used to eliminate fouling. Here usually the continuous phase wets the walls of the channel, and the dispersed phase forms liquid slugs or drops. There is a thin film of the continuous phase between the slug and the wall. The particles are synthesized in the dispersed phase. The segmented flow pattern results in the generation of monodispersed particles [3–6].

Due to the presence of the organic phase, the applications of the nanoparticles synthesized in segmented flows are limited. To overcome this disadvantage, in this work, we study the flow of an aqueous two-phase system (ATPS). ATPS is formed by dissolving either two incompatible polymers or one polymer and inorganic salt. One phase is predominantly rich in one component (typically polyethylene glycol, PEG), and the other phase is rich in the inorganic salt or the other polymer. ATPS provides a mild environment for carrying out the synthesis of biochemical compounds [10]. Hence, metal nanoparticles synthesized in ATPS can be used for biological applications. Krishna et al. developed a framework for simultaneous synthesis and separation of silver nanoparticles in the batch mode [11]. They used an ATPS with PEG as polymer and trisodium citrate dihydrate as the inorganic salt. Both these components act as reducing agents for the reduction of silver nitrate to silver. The concentration of these reducing agents was very high to ensure two-phase formation. Silver nanoparticles were synthesized in one phase, and later the complementary phase was added. The amount of complimentary solution added is such that the resulting mixture forms two phases. The formed nanoparticles get trapped at the interface, which is then separated and analyzed.

The translation of this to continuous synthesis would help in scaling up the synthesis of nanoparticles. Motivated by this, in this work as a first step, we investigate the hydrodynamics of ATPS at small length scales experimentally. We also compare our experimental results with the predictions of numerical simulations.

The flow regime maps for typical organic-aqueous systems have been extensively investigated. These are characterized by significantly high interfacial tensions [12–20]. ATPS systems are characterized by very low interfacial tension values (by at least two orders of magnitude). In this work, we have considered the PEG - citrate ATPS system. One important objective of this work is to see if any new physics arises in an ATPS system. Specifically, we ask the question that the flow regime maps reported for organic-aqueous systems be used for ATPS systems.

The flow regimes are determined by the dimensionless capillary number prevailing in the system. We now compare the dimensionless Capillary number of the continuous phase ($Ca_c$) for an aqueous-organic system represented by castor oil-water system and ATPS at two typical flow rates in Table 1. $Ca_c$ for the oil-water system is low compared to ATPS at both the flow rates for a microfluidic system. The low interfacial tension results in high values of Capillary number for ATPS in a microfluidic system. Consequently, only stratified flow or core annular flow is observed in a microfluidic system depending on the geometry of the system. It is hence difficult to observe segmented flow in a microfluidic channel when using an ATPS. However, if we were to employ a millichannel, $Ca_c$ becomes lower for an ATPS system as seen in Table.1. Hence to observe both slug flow and stratified flow in an ATPS, a milli channel is used in this work. This choice would allow us to study nanoparticle synthesis in both stratified as well as slug flows in the system.

To study nanoparticle synthesis using the system of Krishna et al.[11] in continuous mode, the operating conditions which result in the different flow regimes for an ATPS are the focus of this work. The small length scales result in the narrow size distribution of the particles synthesized and the use of ATPS results in a green process. Numerical simulations are performed for this two-phase system to understand if there are any numerical challenges at the low interfacial tension values of this system. This also helps us to generalize the results obtained and extend them to other ATPS systems.

Table 1. The capillary number of continuous phase for castor oil-water system and ATPS in millichannel and microchannel.

|  | Castor oil- water | | ATPS (PEG -Citrate) | |
|---|---|---|---|---|
|  | $\rho_{oil}$= 935 kg/m$^3$ <br> $\rho_{water}$= 997 kg/m$^3$ <br> $\mu_{oil}$=0.866 kg/m.s <br> $\mu_{water}$=0.894x10$^{-3}$ kg/m.s <br> $\gamma_{ow}$=18.3x10$^{-3}$ N/m <br> Continuous phase= castor oil | | $\rho_{PEG}$= 1078 kg/m$^3$ <br> $\rho_{Citrate}$= 1124 kg/m$^3$ <br> $\mu_{PEG}$=0.02205 kg/m.s <br> $\mu_{Citrate}$=0.002 kg/m.s <br> $\gamma_{ATPS}$=12.5x10$^{-5}$ N/m <br> Continuous phase= PEG | |
|  | $Ca_{c,OW} = \dfrac{\mu_{oil} U_c}{\gamma}$ | | $Ca_{c,ATPS} = \dfrac{\mu_{PEG} U_c}{\gamma}$ | |
|  | $Q_c=Q_d$= 10 µL/min | $Q_c=Q_d$= 100 µL/min | $Q_c=Q_d$= 10 µL/min | $Q_c=Q_d$= 100 µL/min |
| **Millichannel ($d_h$= 1 mm)** | 0.01 | 0.1004 | 0.0374 | 0.3743 |
| **Microchannel ($d_h$= 100 µm)** | 1.0042 | 10.0421 | 3.7433 | 37.4332 |

Slug flow, transition flow, and core annular flows are observed in a milli-fluidic system with ATPS. Slug length characteristics, film thickness were experimentally quantified. These were verified with simulations performed in ANSYS FLUENT.

## 2. Materials and methods:

### 2.1 Materials:

Polyethylene glycol (M.W. = 6000) was purchased from Sisco Research Laboratories Pvt Ltd., India, and trisodium citrate dihydrate ($C_6H_5O_7Na_3.2H_2O$), LR grade was purchased from Avantor Performance Materials, India. Water was used from Evoqua Water Technologies (Pure Lab Ultra). Methylene blue dye stain was purchased from Merck Life Sciences Private Limited, India.

### 2.1.1 Preparation of reagents:

Pure PEG solution (24 wt %) was prepared by dissolving 24 grams of PEG in 76 grams of water. Similarly, pure citrate solution (24 wt %) was prepared. Equal volumes (100 ml) of this pure PEG and pure citrate solutions were mixed with 1-2 mg of methylene blue dye for 1 min in a beaker placed on a magnetic stirrer (Remi 2ML) at 250 rpm. This is transferred to a separating funnel. After 8 hours it attains equilibrium and forms two immiscible phases [11], a PEG rich (top phase) and a citrate rich phase (bottom phase). The composition of these solutions corresponds to those at the end of the tie line. The two phases obtained are used for hydrodynamic studies. These two solutions one is PEG rich and the other is citrate rich are in equilibrium, These two solutions which are at equilibrium are pumped into the microchannel to avoid any mass transfer inside the channel. This would enable us to obtain uniform size slugs. The refractive index of the two solutions is similar and hence it is necessary to add methylene blue to facilitate visualization of the two phases. The PEG-rich solution is blue in color and the citrate-rich solution remains colorless.

## 2.2 Hydrodynamic studies:

A silicone tube of inner diameter 1 mm and length 1 m was used as a milli channel. Silicone tubes of the same diameter were used to connect the syringes via a T-junction to the milli channel inlet. One syringe was filled with a PEG-rich solution and the other with a citrate-rich solution. Two syringe pumps (Harvardpump 11 elite 70–4501) were used to pump the liquids. A schematic of the experimental setup is shown in Figure 1. Images were captured at the imaging region (red dotted line) which is 10 cm away from T-Junction as shown in Figure 1.

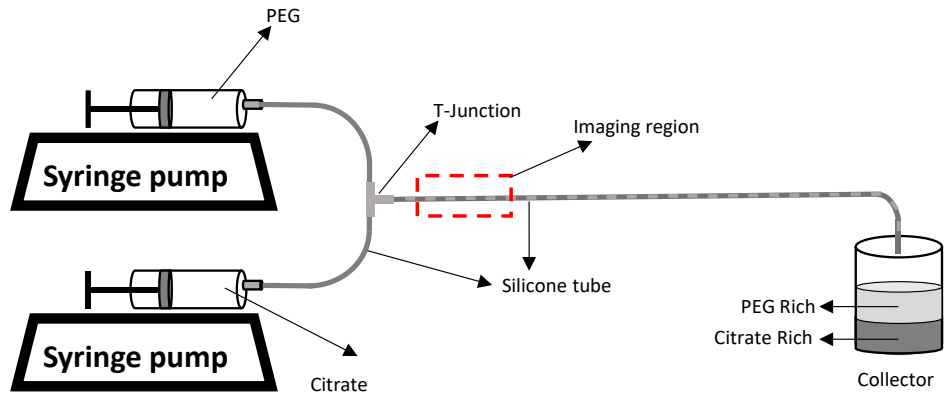

Figure 1. Schematic of the experimental setup used for hydrodynamics of ATPS in milli channel reactor.

The flow rate of each stream was varied from 10 μL/min to 100 μL/min and the flow pattern was observed and recorded. For this ATPS system, the continuous phase was PEG and the dispersed phase was citrate. Images of slugs and jetting regimes were captured by attaching a Skyvik Signi Macro lens (15x) to an iPhone 12 pro max. The images were further processed to estimate slug length, film thickness, and distance between slugs (inter slug distance).

The effect of various dimensionless groups like Reynolds number ($Re$), Capillary number ($Ca$), and Weber number ($We$) on the flow regime maps were analyzed.

### 2.3 Characterizations:

The images captured were analyzed using ImageJ and MATLAB. The viscosity of PEG-rich and citrate-rich solutions were measured using a viscometer (Brookfield LVDV-II+Pro). The density of the phases was measured using a density meter (Rudolph DDM 2909). The refractive index of the two phases was measured by a refractometer (Rudolph Research J357). These values are reported in Table 2 and used in numerical studies.

### 3. Mathematical modeling:

To numerically investigate the hydrodynamic behavior of slug formation in an aqueous two-phase system, simulations were performed in Ansys Fluent 2021. We now describe the details for the numerical simulations

## 3.1 Model geometry

A 3D T-junction milli channel with a circular cross-section of 1 mm diameter and a channel length of 100 mm with two side entrances each of length 2.5 mm was employed as shown in Figure 2. As working fluids, polyethylene glycol (PEG) and trisodium citrate dihydrate are used. The fluid properties of working fluids used in the simulations are listed in Table 2.

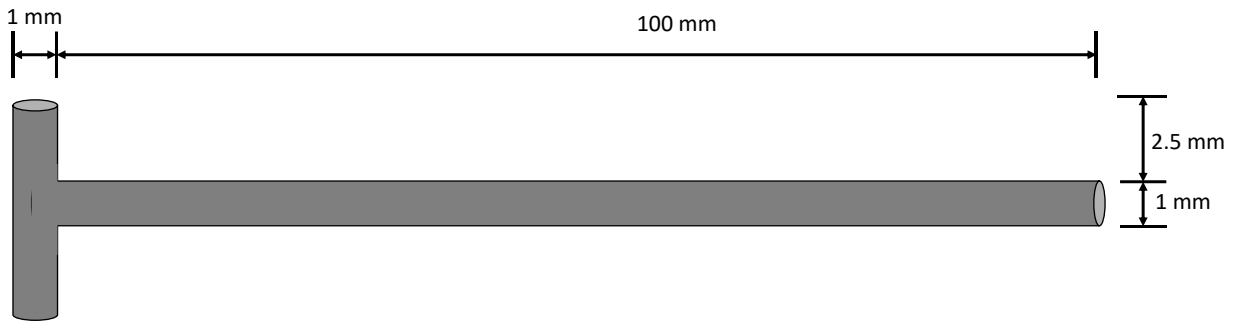

Figure 2. Schematic of milli channel used for numerical simulations.

Table.2 Fluids property measured and used in the simulation

| Property | Value |
| --- | --- |
| Density of PEG ($\rho_c$) | 1078 kg/m$^3$ |
| Density of Citrate ($\rho_d$) | 1124 kg/m$^3$ |
| Viscosity of PEG ($\mu_c$) | 0.02205 kg/m.s |
| Viscosity of Citrate ($\mu_d$) | 0.002 kg/m.s |
| Refractive index of PEG | 1.3608 |
| Refractive index of Citrate | 1.3788 |

## 3.2 Governing equations

The classical volume of fluid (VOF) technique was used to capture the liquid-liquid interface. The fluids in this approach share a single set of continuity and momentum equations, and the volume fraction of each fluid was monitored in each computing cell across the domain. Each computational cell is occupied by two immiscible liquids. The volume fraction of the first fluid is denoted by $\alpha$ in a computational cell. If $\alpha = 1$ the computational cell has only fluid 1 while if $\alpha = 0$ the computational cell has only fluid 2. For $0 < \alpha < 1$ the cell is occupied by both fluids and the cell has an interface.

The continuity and momentum equations used are

$$\frac{\partial \rho}{\partial t} + \nabla \cdot (\rho \vec{v}) = 0 \tag{1}$$

$$\frac{\partial \rho \vec{v}}{\partial t} + \nabla \cdot (\rho \vec{v} \vec{v}) = -\nabla p + \nabla \cdot [\mu (\nabla \vec{v} + \nabla \vec{v}^T)] + \vec{F} \tag{2}$$

Where $\rho$ is the density, $\mu$ is the viscosity, $\vec{v}$ is the velocity vector, $p$ is pressure, and $\vec{F}$ is the surface tension force. The fluid characteristic density($\rho$) and viscosity($\mu$) of a two-phase mixture are defined as

$$\rho = \alpha \rho_1 + (1 - \alpha) \rho_2 \tag{3}$$

$$\mu = \alpha \mu_1 + (1 - \alpha) \mu_2 \tag{4}$$

The evolution of volume fraction($\alpha$) used to track the interface between the two phases is governed by

$$\frac{\partial \alpha}{\partial t} + \vec{v} \cdot \nabla \alpha = 0 \tag{5}$$

To describe surface tension effects in micro and millichannels, a volumetric source term ($\vec{F}$) is incorporated into the momentum equation. For this, we adopt the continuum surface model proposed by Brackbill et al. [21].

$$\vec{F} = \sigma \frac{\rho \kappa_G \nabla \alpha_G}{\frac{1}{2}(\rho_G + \rho_L)} \tag{6}$$

Here $\kappa_G$ is the curvature determined from the divergence of the unit surface normal, while n is the surface normal vector.

$$\kappa_G = -\nabla \cdot \hat{n} \tag{7}$$

$$\hat{n} = \frac{\nabla \alpha}{|\nabla \alpha|} \tag{8}$$

**3.3 Numerical simulation conditions**

A 3D T-junction microchannel geometry was built and meshed in Design Modular (ANSYS 2021 R1). This was imported for post-processing to analyze features of the different flow patterns observed. The simulations are carried out in ANSYS 2021 R1 FLUENT software with double-precision solvers. To achieve optimal mesh spacing, grid independence tests were performed. The computational geometry used has 400132 cells. In all simulations, the flow conditions were laminar, and the Capillary numbers were in the range $10^{-3}$ - $10^{-1}$. The simulations were run on a PC with an Intel® Core (TM) i7-7700 CPU @3.60GHz and 32GB RAM. For the continuity and momentum equation, the scaled convergence criterion used was $10^{-3}$.

The Piecewise Linear Interface Construction (PLIC) algorithm was used for interface reconstruction, while the Pressure Staggering Option (PRESTO) approach was used for pressure interpolation. The Pressure-Implicit with Staggering Option (PISO) algorithm and the second-order upwind technique were employed for the pressure-velocity coupling and momentum equation, respectively.

The channel is initially filled with a PEG-rich phase. The PEG-rich and Citrate-rich phases were introduced into the channel at a constant velocity at time zero. At the exit, the flow was considered fully developed. At the channel's walls, no-slip boundary conditions are used. The interfacial tension was set to $1.25 \times 10^{-4}$ N/m. This value was chosen as it was found to fit well with experimental data. The wall contact angle of the PEG was set to 0.

## 4. Results and discussion:

We first discuss the experimentally observed flow regimes for different operating conditions. We then discuss the quantification of the flow characteristics and compare the results with predictions from numerical simulations. We also check the validity of the hydrodynamic correlations developed for organic-aqueous systems.

**4.1. Flow regime of ATPS:**

The physical properties of fluids such as interfacial tension ($\gamma$), the viscosity of continuous phase ($\mu_c$), dispersed phase ($\mu_d$) and density of continuous phase ($\rho_c$) and dispersed phase ($\rho_d$) characterize the generation of slugs or drops. In this system, the continuous phase was PEG rich and the dispersed phase was citrate rich. The effect of these variables is captured through dimensionless numbers.

The Capillary number *Ca* for each phase was evaluated for this system using:

$$Ca_c = \frac{\mu_c U_c}{\gamma} \tag{9}$$

$$Ca_d = \frac{\mu_d U_d}{\gamma} \tag{10}$$

Here $U_c$ and $U_d$ are superficial velocities of the continuous phase and dispersed phase respectively. The dimensionless *Ca* plays an important role in slug/droplet formation.

The relative ratio of fluid inertia to capillary pressure is determined by the dimensionless Weber number (*We*).

$$We_c = \frac{\rho_c d U_c^2}{\gamma} \tag{11}$$

$$We_d = \frac{\rho_d d U_d^2}{\gamma} \tag{12}$$

*We* represent the ratio of inertial forces to interfacial tension forces. *Ca* and *We* help us to predict the onset of jetting. For example, slugs are obtained when the instabilities in two-phase stratified flow get amplified. This can occur at low flow speeds or when the ratio of viscous forces to interfacial tension i.e., *Ca* is low.

The flow regime obtained for various flow rates is shown in Figure 3. Slug flow is observed when the flow rate of the dispersed phase (Citrate rich) is lower than 20 µL/min for all values of the continuous phase (PEG rich) flow rate. With an increase in the flow rate of the dispersed phase, a transition region is observed where jetting takes place. Here the two fluids flow as a core annular flow near the entrance and this eventually breaks into slugs. At an even higher flow rate (above 80 µL/min) of the dispersed phase, core annular flow is observed throughout the channel for low values of the flow rate of the PEG-rich phase. For flow rates above 100 µL/min for the dispersed phase core-annular flow is observed for all flow rates of the continuous phase.

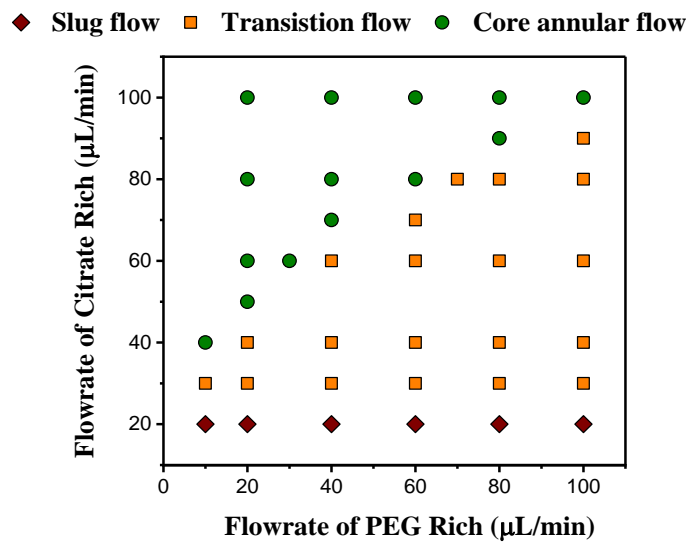

Figure 3. Flow regime map for PEG -citrate ATPS in a milli channel system.

To compare this map of the aqueous two-phase system ATPS with the aqueous-organic system (water-castor oil)[12], we depict the flow regime maps in terms of dimensionless numbers. For ATPS in the milli channel, three regimes are observed (Slug, transition, and core annular) as shown in Figure 3 and Figure 4. Flow pattern maps based on the Reynolds number (*Re*), Capillary number (*Ca*), and Weber number (*We*) of each phase were plotted for ATPS. These maps show ranges of dimensionless parameters that determine the balance of forces in each flow regime.

The ratio of the inertial to viscous forces of the citrate-rich phase ($Re_d$) varies significantly for the range of flow conditions studied. When $Re_d$ is much lesser than 0.3 ($Re_d < 0.3$) inertia is negligible compared to viscous forces, and this resulted in slug flow. For $Re_d > 0.3$, inertia starts to affect the flow. When $Re_d$ is near 0.7, then inertia becomes strong enough to establish core annular flow. So for $Re_d \geq 0.7$, core annular flow is observed.

In the ($We_c$, $We_d$) plane, inertial effects in the citrate-rich phase become comparable with interfacial tension ($We_d > 0.1$). In these regions, parallel flow (core annular) and transition flow is observed. $Ca_c$ and $Ca_d$ was evaluated and the flow regime map was plotted in the $Ca_c$ -$Ca_d$ plane. In the ($Ca_c$, $Ca_d$) plane, viscous effects in the PEG phase are comparable with interfacial tension for ($Ca_c > 0.3$). Here slug flow or transition flow is observed. This corresponds to a flow rate of the dispersed phase (citrate) which is less than 100 μL/min. These trends are qualitatively similar to those reported for the castor oil-water system. A comparison of the transition values in terms of dimensionless numbers for the two systems is shown in Table 3.

Table 3. Comparison of flow regime from the organic-aqueous system and aqueous-aqueous system in terms of dimensionless numbers.

| Castor oil-water system | | ATPS (PEG-Citrate system) | |
|---|---|---|---|
| $We_d > 0.1$ | Parallel flow | $We_d > 0.1$ | Core annular or transition |
| $Ca_c > 0.1$ | Droplet flow | $Ca_c > 0.3$ | Slug or transition |
| $Re_d < 0.1$ | Slug | $Re_d < 0.3$ | Slug |
| $0.1 < Re_d < 1$ | Transition from slug to core annular | $Re_d > 0.3$ | Transition |
| $Re_d = 1$ | Core annular | $Re_d \geq 0.7$ | Core annular |

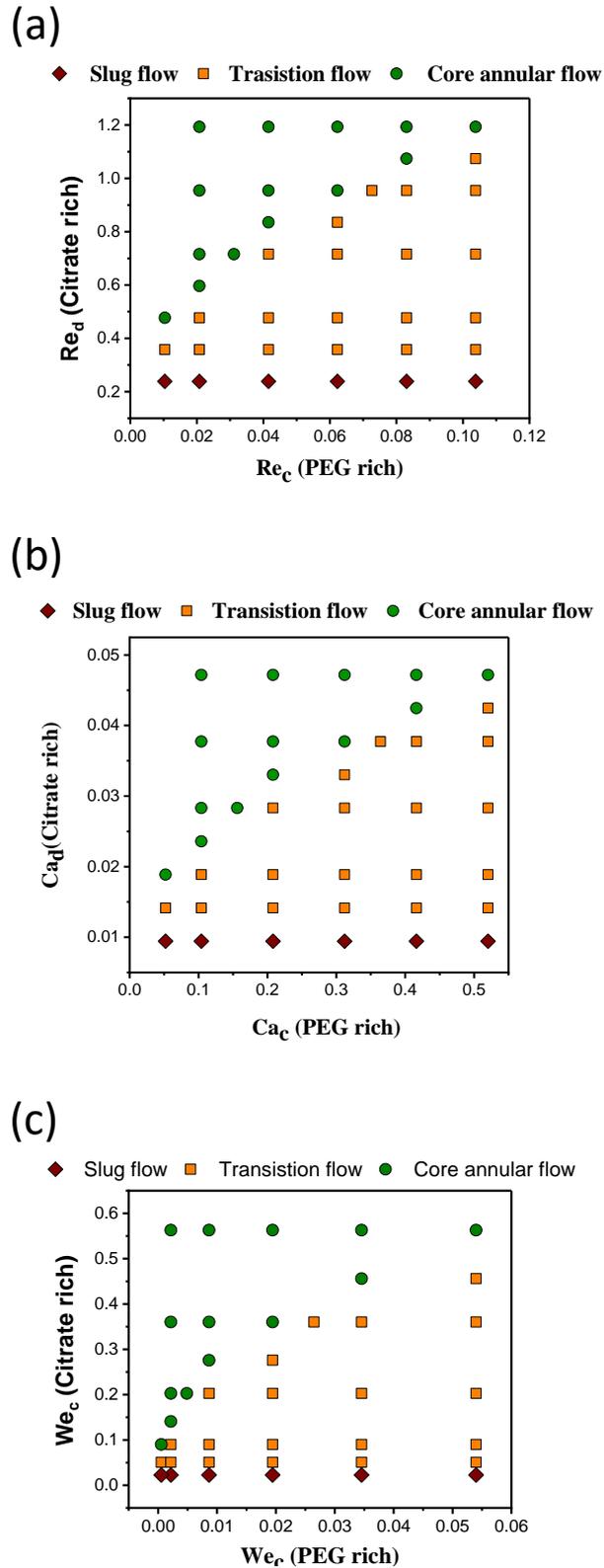

Figure 4. Flow regime map in terms of *Re*, *Ca*, and *We* of the continuous phase (PEG) and the dispersed phase (citrate).

## 4.2. Hydrodynamics of ATPS with an equal flow rate of each phase:

To quantitatively describe the slug flow observed, experiments were done with equal flow rates of each phase. Here we obtained the slug length and the film thickness for the different flow conditions studied. The total flow rate was varied from 20 µL/min to 200 µL/min. This corresponds to a range of $Ca_c$ from 0.037 to 0.37, and $Ca_d$ from 0.0034 to 0.034.

For the PEG-Citrate ATPS system with equal flow rates, we found that there was slug formation for $Ca_c \leq 0.208$ and jet formation for $0.28 < Ca_c \leq 0.83$ For $Ca_c > 0.37$ only core annular flow is observed.

Further, images were analyzed for equal flow rates to determine slug length, slug diameter, and inter slug distance. Images captured for the PEG-rich phase with methylene blue and citrate-rich system are shown in Figure 5.

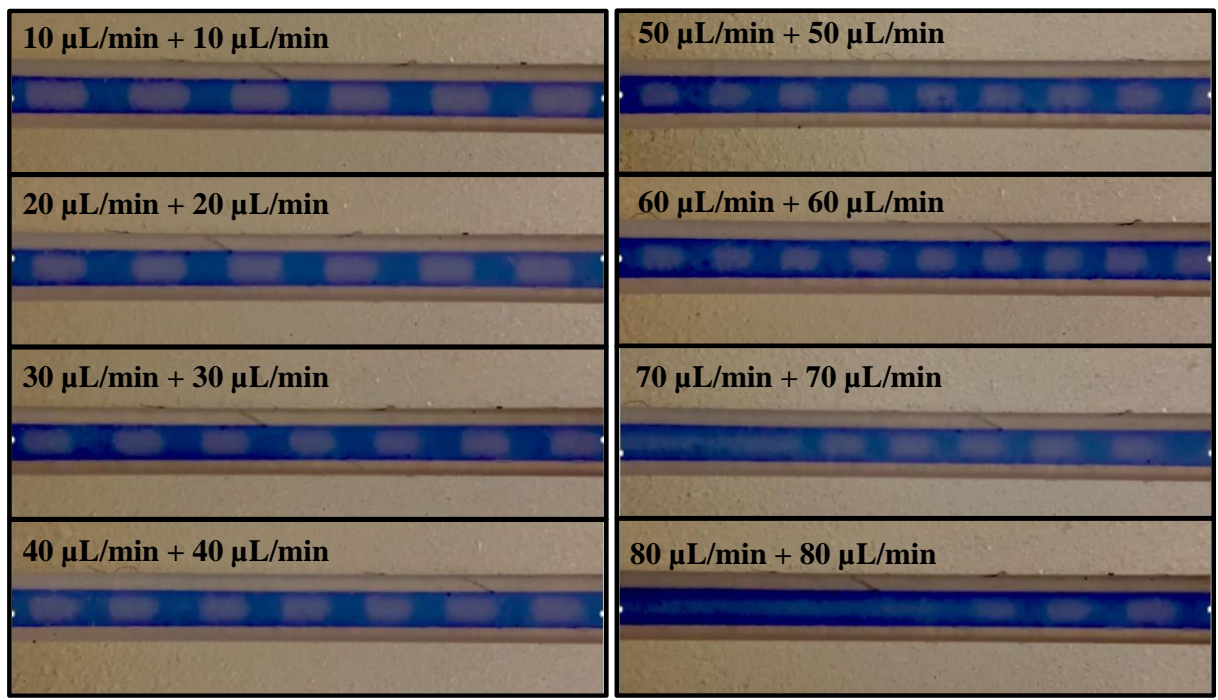

Figure 5. Images of slugs taken at various flow rates of the two phases for equal flow rates.

The captured image was post-processed to get the actual slug length. Analysis of images was performed using ImageJ software and MATLAB. In ImageJ, the size of the slug, the distance between slugs, and film thickness were analyzed manually. To verify this, RGB data of the image was extracted in MATLAB and converted to greyscale. The threshold value for the greyscale was extracted in MATLAB and this value was used for converting greyscale image to black and white image in MATLAB as shown in Figure 6. The centerline data of the resulting black and white image was a logic matrix (which takes the value 1 for white and 0 for black) as shown in Figure 6. This data (pixel) was used for analyzing the length of the slug. Slug size was obtained by taking an average of 5 slugs. The size of the slugs measured using ImageJ and MATLAB is shown in Figure 7. The sizes obtained using both methods are in good agreement. Slug diameter and inter slug distance were similarly analyzed in ImageJ and this is shown in Figure 8. It is observed that with an increase in flow rate of the two phases the slug length, slug diameter, and inter slug distance decreases.

When the total flow rate is above 40 μL/min jetting takes place. The inner fluid moves as a jet before it breaks into slugs. The length at which the jet breaks into slugs is called the jetting length. The change in jetting length with total flowrate is shown in Figure 9. In these experiments, the jetting length was found to vary periodically. This is shown for combined flowrate of 160 μl/min in the video V1 in the supplementary material. The low and high values of jetting length correspond to the minima and maxima observed in the jetting length. In Figure 9, the average of three experiments is reported. The dashed line and solid line are the best fit curve for minimum and maximum observed in the jetting length. These fluctuations can be eliminated by using suitable flow dampeners in the system to avoid fluctuations in the inlet velocity to the channel.

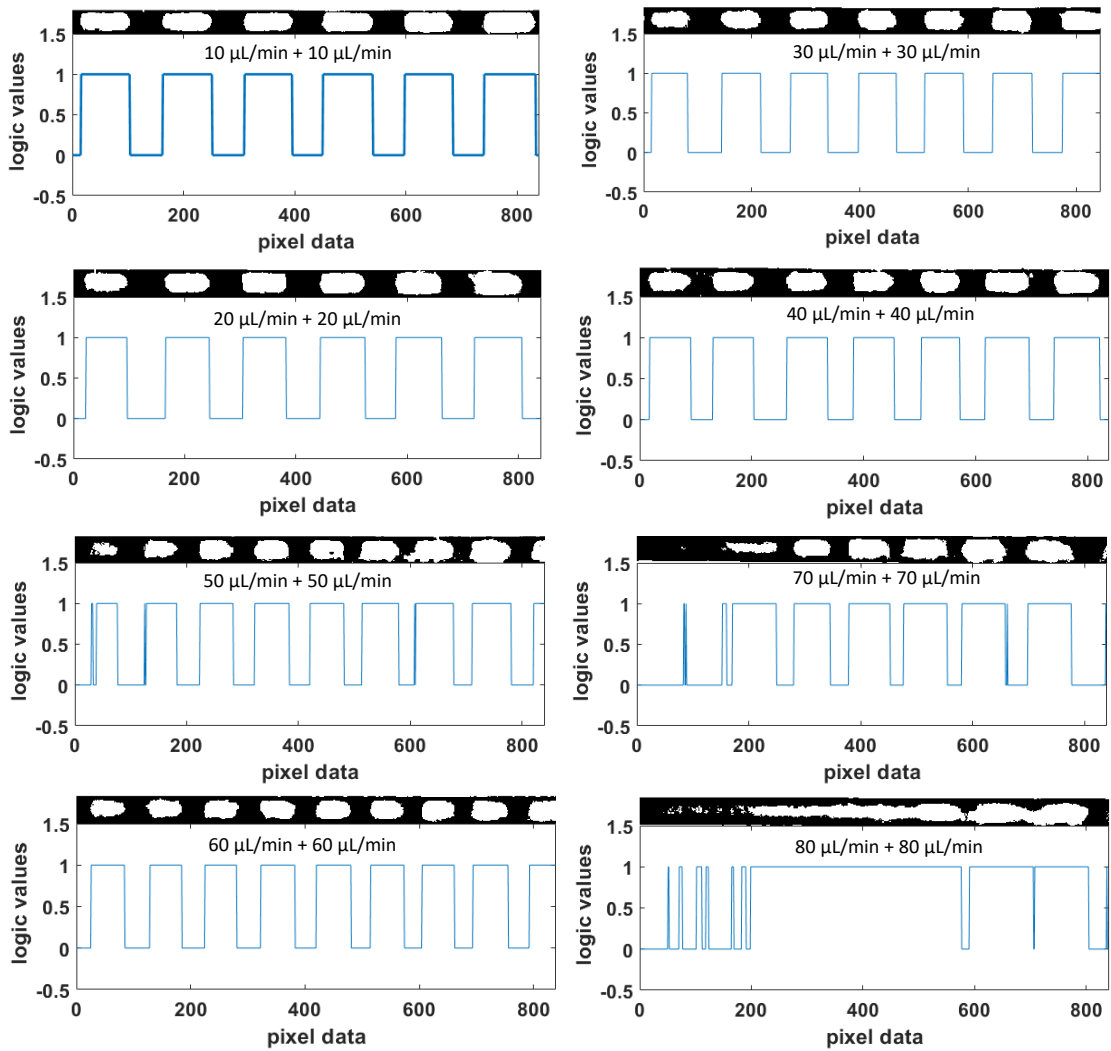

Figure 6. Black and white images of slugs obtained after postprocessing the actual experimental image. Also shown is the logical output data along the center horizontal line.

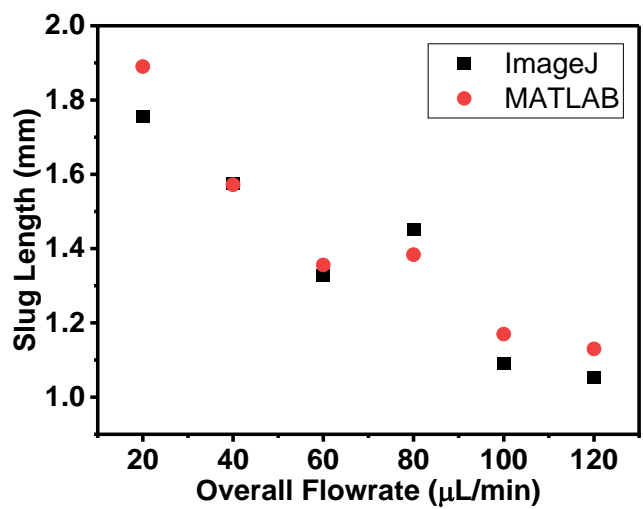

Figure 7. Comparison of slug lengths obtained using ImageJ software and MATLAB post-processing.

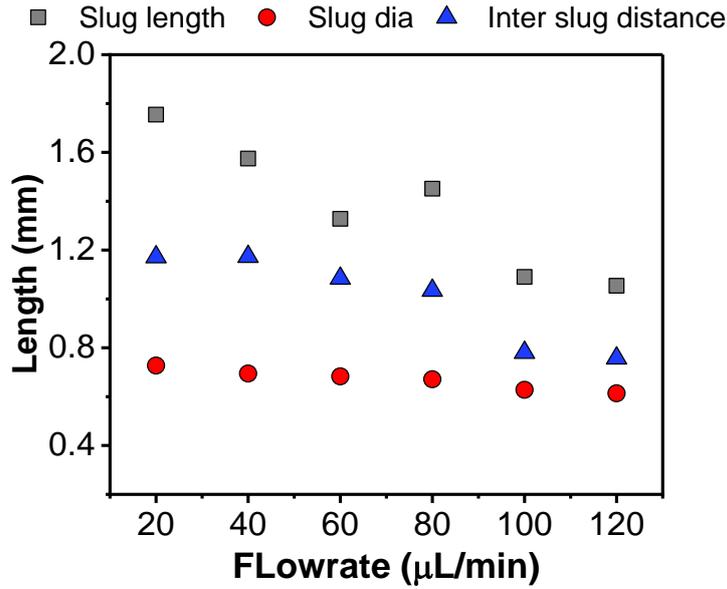

Figure 8. Slug length, diameter, and distance between slugs were measured using ImageJ analysis.

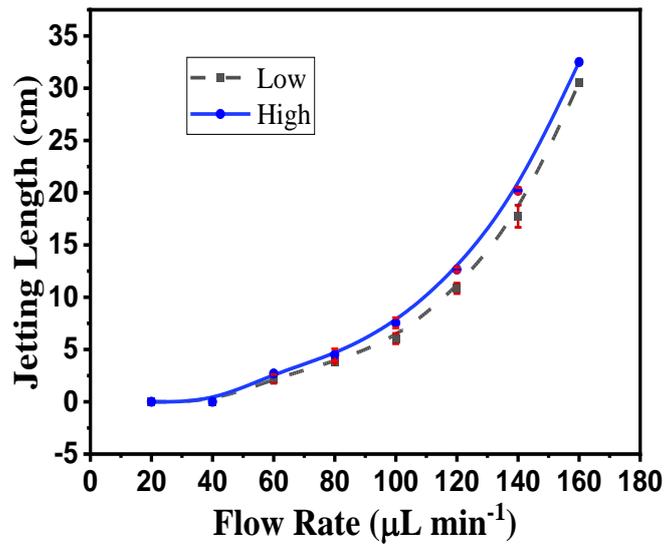

Figure 9. Variation in jetting length with change in flow rate. Here the jetting fluctuates at a specific flow rate. The dashed line represents the minimum value and the solid represents the maximum value between which the jetting length fluctuates.

### 4.3. Numerical simulations:

Simulations were performed for various interfacial tension values and the slug lengths predicted were compared with experimental data. It was observed that for an interfacial tension value of $1.25 \times 10^{-4}$ N/m the predicted values of slug length were matching well with experimental data. The images of the slugs obtained from simulations are shown in Figure 10. These are the images captured at a distance of 10 cm from the T-junction to enable comparison with the experimental data. The slug length of simulation results is compared with the experimental values in Figure 11. It can be seen that the simulation results are in good agreement with the experimental data. Simulation results also show the formation of jets at a higher flow rate.

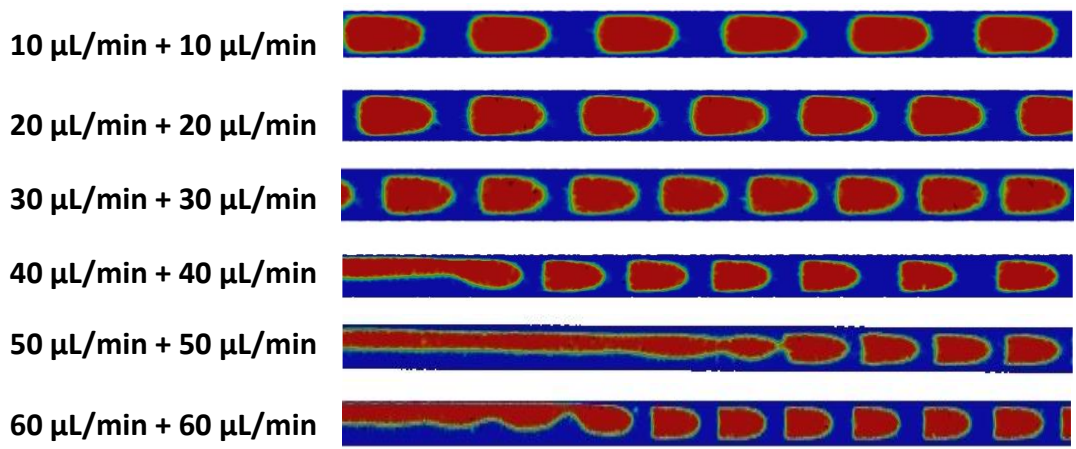

Figure 10. Images of the slugs obtained from simulation and the jetting observed at high flow rates.

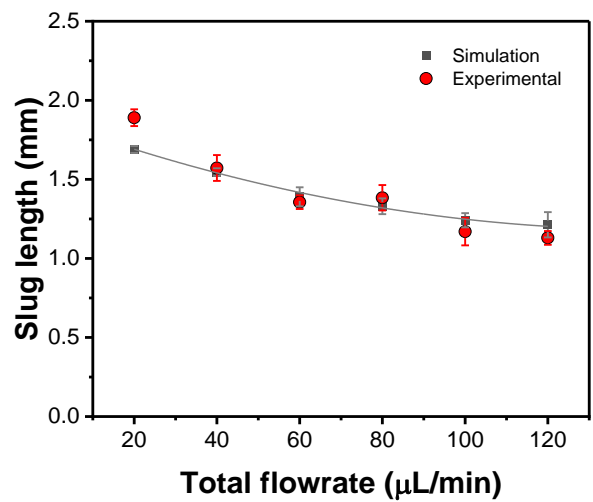

Figure 11. Comparison of slug length of simulation predictions with experimental data (MATLAB Analysis).

### 4.4. Film thickness comparision:

The slug flow regime is characterized by the film thickness. A thin film of continuous phase separates the dispersed phase from the walls. The experimental images and simulation predictions were analyzed using ImageJ software and compared with the correlation proposed in the literature [10,22]. The correlation for the thickness of the film in terms of $Ca_o$, $We_o$, $Re_o$ is:

$$\frac{\delta}{d} = \frac{0.670(Ca_o)^{\frac{2}{3}}}{1 + 3.13(Ca_0)^{\frac{2}{3}} + 0.504(Ca_0)^{0.672}(Re_o)^{0.589} - 0.352(We_o)^{0.629}} \qquad (13)$$

Here $\delta$ is film thickness. For this, the Capillary number ($Ca_o$) is defined as:

$$Ca_o = \frac{\mu_c U}{\gamma} \frac{\mu_c}{\mu_d} \qquad (14)$$

where $U$ is the total velocity.

Weber number ($We_o$) is defined as:

$$We_o = \frac{\rho_c d U^2}{\gamma} \qquad (15)$$

where $d$ is the hydraulic diameter.

Reynolds ($Re$) is defined as:

$$Re_o = \frac{We_o}{Ca_o} = dU\rho_c \frac{\mu_d}{\mu_c^2} \qquad (16)$$

If the flow rates of both the phases are equal it is expected that the film thickness will increase with the increase in total flowrate [10].

The film thickness measured experimentally is compared with the simulation predictions and the correlation in Figure 12. It can be seen that the simulation predictions and experimental measurements agree well with the hydrodynamic correlation. The film thickness increases with an increase in the total flow rate.

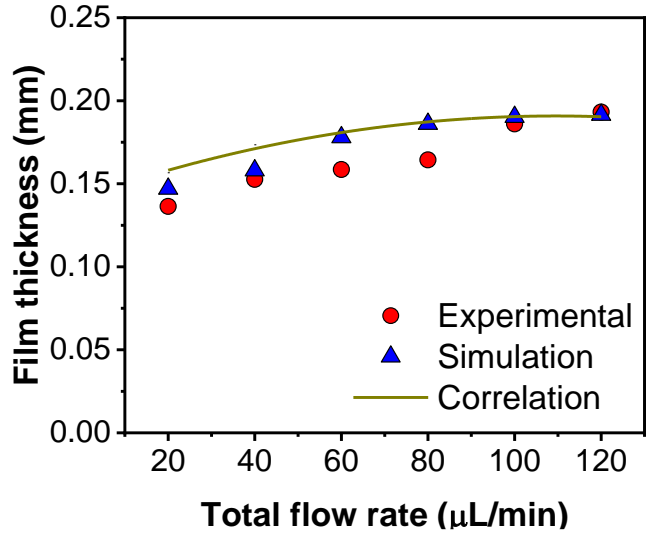

Figure 12. Comparison of the film thickness of experimental data with simulation results and correlation.

**4.5. Core annular flow quantification:**

For core annular flows, Vir et al.[23] developed a relation between holdup and flowrate fraction. The holdup of the core fluid in core-annular flow ($\alpha_c$) is defined as volume occupied by the core fluid to the total volume of the channel,

$$\alpha_c = \frac{R_i^2}{R_o^2} \qquad (17)$$

Here $R_i$ is the radius of the core and $R_o$ is the channel radius.

The viscosity ratio is defined as the ratio of viscosity of the core fluid to that of the annular fluid

$$\mu = \frac{\mu_c}{\mu_a} \qquad (18)$$

For the PEG-Citrate ATPS system, the core fluid is citrate rich, and the annular fluid is PEG rich and this ratio is 0.0907 (Table 2).

The flowrate fraction $\emptyset$ of the core fluid in core-annular flow ($\emptyset$) is related to hold up through

$$\emptyset = \frac{[\alpha_c^2 - 2\mu\alpha_c^2 + 2\mu\alpha_c]}{[\alpha_c^2 - \mu\alpha_c^2 + \mu]} \tag{19}$$

For equal flow rates of the PEG and citrate phase, the flowrate fraction $\emptyset$ is 0.5.

The holdup evaluated with the above equation is 0.1829. Experimentally the core radius was measured using ImageJ as 0.424 mm. The holdup value measured was 0.179 which agres well with the theoretically predicted value.

## 5. Conclusions:

In this paper, we analyzed the flow behavior of an Aqueous Two-Phase System in a millichannel. For the experiments, a PEG 6000 and citrate system is chosen. This choice is motivated by the fact that both these compounds act as reducing agents in nanoparticles synthesis. ATPS systems are characterized by low interfacial tension values. The capillary number of continuous phase for ATPS is consequently high. As a result, we obtain parallel flow in the microchannel at typical flow rates used in the experiments. To obtain slug flow we would have to go to a very low flow rate. This would result in a low throughput of the system. To overcome this problem and obtain segmented flow at a reasonable throughput we propose that millichannels has to be employed. This allows us to observe both stratified flow and dispersed phase flow in the system. Using a millichannel also helps us operate at a low-pressure drop. The two fluids which are pumped into the millichannel are at equilibrium. This eliminates any mass transfer effects from the PEG-rich phase to the citrate-rich phase. Consequently, the results obtained in this work are only governed by hydrodynamics.

The flow regime map of ATPS was similar to that observed in an aqueous-organic system. The transition values from one flow pattern to another occur at dimensionless numbers whose order of magnitude is similar.

The flow regime maps help us identify operating conditions under which the different flow patterns are observed. This allows the engineer to choose conditions which are favorable for their particular application.

Simulation results were obtained for different values of interfacial tension. These predictions were matched with the experimental results to obtain the value of the interfacial tension for this system. An interfacial tension value of $1.25 \times 10^{-4}$ N/m was found to be optimal i.e simulation predictions matched well with experimental data. The low values of interfacial tension also did not yield any numerical challenges in the simulation. This we attribute to the employment of the Brackbrill approach for simulating the system.

Finally, the correlations developed in the literature for aqueous organic systems seem to be valid for the ATPS system. The results presented in this work will help develop new applications using ATPS in millichannel systems. The use of ATPS will help in developing a green process for novel applications.